\documentclass[conference]{IEEEtran}
\usepackage{pifont}
\usepackage{xcolor}
\usepackage{amsmath,amsfonts}
\usepackage{array}
\usepackage[caption=false,font=normalsize,labelfont=sf,textfont=sf]{subfig}
\usepackage{textcomp}
\usepackage{stfloats}
\usepackage{url}
\usepackage{enumitem}
\usepackage{verbatim}
\usepackage{graphicx}
\usepackage{algorithm}
\usepackage{algpseudocode}

\usepackage{booktabs} 
\usepackage{stackengine}
\usepackage{float}
\usepackage{tikz}
\usepackage{xcolor}
\usepackage{wasysym}
\usepackage{colortbl}
\usepackage{soul}
\usepackage{multirow}
\usepackage{ragged2e}
\usepackage{hyperref}
\usepackage{comment}
\newcommand\encircle[1]{%
\tikz[baseline=(X.base)]
 \node (X) [draw, scale=0.75, shape=circle, inner sep=0, fill=black, text=white, minimum size=0em] {\strut #1};}

 \newcommand{\cmark}{\ding{51}}%
\newcommand{\xmark}{\ding{55}}%

\ifCLASSINFOpdf
\else
\fi
\begin{document}
\title{SPICEPilot: Navigating SPICE Code Generation and Simulation with AI Guidance \vspace{-0.5em}
}


\author{Deepak Vungarala, Sakila Alam, 
      Arnob Ghosh, Shaahin~Angizi \\
\small

Department of Electrical and Computer Engineering, New Jersey Institute of Technology, Newark, NJ, USA\\
E-mail: \{dv336, sa3229, arnob.ghosh, shaahin.angizi\}@njit.edu\\\vspace{-1.5em}
\vspace{-1em}}

\maketitle

\begin{abstract}
Large Language Models (LLMs) have shown great potential in automating code generation; however, their ability to generate accurate circuit-level SPICE code remains limited due to a lack of hardware-specific knowledge. In this paper, we analyze and identify the typical limitations of existing LLMs in SPICE code generation. To address these limitations, we present SPICEPilot—a novel Python-based dataset generated using PySpice, along with its accompanying framework. This marks a significant step forward in automating SPICE code generation across various circuit configurations. Our framework automates the creation of SPICE simulation scripts, introduces standardized benchmarking metrics to evaluate LLM's ability for circuit generation, and outlines a roadmap for integrating LLMs into the hardware design process. 
SPICEPilot is open-sourced under the permissive MIT license at \href{https://github.com/ACADLab/SPICEPilot.git}{https://github.com/ACADLab/SPICEPilot.git}.
\end{abstract}

\IEEEpeerreviewmaketitle

\section{Introduction}\label{sec:introduction}
The escalating complexity of modern software and the rapid advancements in hardware technologies present significant challenges in developing innovative circuit solutions. As software systems grow more intricate, the hardware that supports them must also evolve, leading to an intertwined cycle of increasing complexity in both domains. Traditional methods of circuit design and simulation struggle to keep pace with these developments, necessitating new approaches that can efficiently handle the growing demands \cite{zhang2024mgverilog, lu2023rtllm, lai2024analogcoder, liu2024layoutcopilot}. Recently, Large Language Models (LLMs) or Language Models (LMs) have demonstrated remarkable capabilities in generating Python code, offering potential avenues for automating aspects of software development. However, their application in hardware design remains limited due to a lack of inherent hardware domain knowledge. This gap poses a significant obstacle, as hardware design requires a deep understanding of architectural, micro-architectural, and logic levels as well as electronic components, circuit behaviors, and simulation processes that LLMs are not typically trained on \cite{zhang2024mgverilog,lai2024analogcoder}. 

LLMs have recently shown promising solutions for generating digital and micro-architectural designs, e.g.,  MEV-LLM \cite{nadimi2024multi} proposes multi-expert LLM architecture for Verilog code generation. RTLLM \cite{lu2023rtllm}, GPT4AIGChip \cite{10323953}, and SA-DS \cite{vungarala2024sa} enhance design efficiency, showcasing LLMs’ ability to manage complex design tasks and broaden access to AI accelerator design. Nevertheless, the application of LLM in analog circuit design has been limited to a few works. To the best of our knowledge, AnalogCoder \cite{lai2024analogcoder} is among the first Analog circuit generators.

The research presented here opens up a vast array of intriguing research questions that are yet to be explored. Our aim is to address the most pressing of these and propose a structured path for future works. Key questions include:
(\textit{RQ1}): How reliable are LLMs in the context of analog circuit design, and what are their foundational limitations in this domain?
(\textit{RQ2}): What steps are required to develop a specialized LLM tailored specifically for analog circuit design?
(\textit{RQ3}): How can the challenge of data scarcity be addressed in the niche field of analog circuit design?
(\textit{RQ4}): Are there methodologies that would enable LLMs to autonomously generate or enhance datasets needed for analog circuit design?
(\textit{RQ5}): How can LLMs be equipped with logical reasoning capabilities specific to circuit design to ensure effective interpretation and solution of hardware-specific problems?
(\textit{RQ6}): What new metrics and benchmarks should be established to accurately assess the performance, accuracy, and reliability of LLMs in executing hardware design tasks?
These research questions form the basis of our investigation, with the ultimate goal of advancing LLM-driven hardware design solutions.

In this paper, we propose a novel framework that leverages the strengths of LLMs in Python code generation to assist in analog circuit design creation and simulation. By utilizing PySpice\footnote{PySpice is an open source Python module which provides a Python interface to the Ngspice and Xyce circuit simulators.}, a Python library for SPICE simulation, we generate a comprehensive dataset of Python-based SPICE codes that correspond to various transistor models and circuit configurations. This approach effectively bridges the gap between software and hardware domains, enabling the use of LLMs to facilitate SPICE simulations and accelerate the design process.
Moreover, we introduce standardized benchmarking criteria to evaluate the performance and accuracy of the generated circuits. This standardization is crucial for comparing different designs and ensuring that the innovations meet the required specifications and industry standards. Our framework lays the groundwork for future research directions, highlighting the potential for integrating LLMs more deeply into hardware design workflows and paving the way for automated, intelligent circuit generation and optimization.

This research addresses many of the key questions outlined earlier, as we explore the capabilities of LLM's in the analog domain. Noteworthy contributions include: 
\begin{itemize}
    \item We evaluate LLM performance in SPICE code generation, where both open-source and proprietary models are analyzed to examine their ability to generate fundamental circuits;
    \item We propose a framework to overcome data scarcity by utilizing LLMs to generate open-source datasets for analog circuits. A reliable benchmarking metric is introduced to assess the performance of LLMs in hardware design; and
    \item We present a comprehensive road map, outlining potential future advancements to further optimize LLMs for analog circuit design.
\end{itemize}

\begin{table*}[t]
\centering
\caption{Comparison of the Selected LLM-based HDL/HLS generators.} \vspace{-1em}
\scalebox{0.83}{
\begin{tabular}{|l|c|c|c|c|c|c|c|c|c|c|c|c|c|c|c|}
\hline
\rowcolor[HTML]{C0C0C0} 
\multicolumn{1}{|c|}{\cellcolor[HTML]{C0C0C0}\textbf{Property}} & \textbf{Ours} & \textbf{ \cite{thakur2023verigen}} & \textbf{ \cite{10323953}} & \textbf{ \cite{chang2023chipgpt}} & \textbf{ \cite{blocklove2023chip}} & \textbf{ \cite{thakur2023autochip}} & \textbf{\cite{ma2024verilogreader}} & \textbf{\cite{fang2024assertllm}} & \textbf{\cite{liu2024verilogeval}} & \textbf{\cite{zhang2024mg}} & \textbf{\cite{lu2023rtllm}} & \textbf{\cite{lai2024analogcoder}} &\textbf{\cite{liu2024ampagent}} &\textbf{\cite{liu2024layoutcopilot}} \\ 
Function & Analog &Verilog &AI Accel. &Verilog &Verilog &Verilog& Hardware & Hardware &Verilog & NA$^\dagger$ &RTL &Spice &Schematic &Layout
\\& ckt. & Gen. & Gen. & Gen. & Gen. & Gen. & Verf. & Verf. & Gen. & & Gen.& Gen. & Desg. & Desg. 
 \\ \hline
Dataset                             & \cmark           & \cmark(Verilog)       & \xmark              & NA            & NA            & NA  & 
\xmark & \xmark    & \cmark & \cmark & \cmark(Verilog) &\xmark & \xmark & \xmark  \\ \hline
Output format                                & Python            & Verilog            & HLS             & Verilog & Verilog             & Verilog     & Verilog &HDL & Verilog & Verilog & Verilog &Python & Schematic &Layout    \\ \hline
Auto. Verif.   & \cmark              & \xmark            & \xmark              & \xmark            & \xmark             & \cmark   & \cmark & \xmark & \cmark  & \xmark  & \cmark & \xmark & \xmark  & \xmark  \\ \hline
Human in Loop                           & Low     & Medium           & Medium             & Medium           & High            & Low  & Low & Low & Low & Low & Low    & High & Low& Low  \\ \hline
Fine tuning   & \xmark & \cmark  & \cmark & \xmark & \xmark & \xmark & \xmark & \xmark & \xmark & \cmark & \xmark &\xmark & \xmark & \xmark  \\ \hline
\end{tabular}
}
\label{analysis}
\tiny
$^*$A user interface featuring Prompt template generation for the input of LLM.
$^\dagger$ Not applicable.
\vspace{-1.2em}
\end{table*}

 \section{Background}
 \textbf{LLM for Hardware Design.} LLMs show promise in generating Hardware Description Language (HDL) and High-Level Synthesis (HLS) code. 
VeriGen \cite{thakur2023verigen} and ChatEDA \cite{he2023chateda} refine hardware design workflows, automating the RTL to GDSII process with fine-tuned LLMs. AssertLLM incorporates three customized LLM and finally generate multiple system verilog assertions each performing different functionalities \cite{fang2024assertllm}. ChipGPT \cite{chang2023chipgpt} and Autochip \cite{thakur2023autochip} integrate LLMs to generate and optimize hardware designs, with Autochip producing precise Verilog code through simulation feedback. MG-Verilog \cite{zhang2024mgverilog} created a hardware dataset with over 11,000 verilog code. Chip-Chat \cite{blocklove2023chip} demonstrates interactive LLMs like ChatGPT-4 in accelerating design space exploration. MEV-LLM \cite{nadimi2024multi} proposes multi-expert LLM architecture for Verilog code generation.
RTLLM \cite{lu2023rtllm} and GPT4AIGChip \cite{10323953} enhance design efficiency, showcasing LLMs’ ability to manage complex design tasks and broaden access to AI accelerator design. In VerilogReader \cite{verilog_reader_2024} the LLM accurately grasp the code logic and generate stimuli to reach the unexplored code branches. To the best of our knowledge, GPT4AIGChip \cite{10323953} and SA-DS \cite{vungarala2024sa} are a few initial works focus on an extensive framework specifically aimed at the generation of domain-specific AI accelerator designs where SA-DS focus on creating a dataset in HLS and employ fine-tuning free methods such as single-shot and multi-shot inputs to LLM. LLMCompass  \cite{zhang2024llmcompass} is able to describe and evaluate different hardware design. 
However, the \textit{absence of prompt optimization, tailored datasets, model fine-tuning, and LLM hallucination} pose a barrier to fully harnessing the potential of LLMs in such frameworks \cite{he2023chateda,vungarala2024sa}. This limitation confines their application to standard LLMs without fine-tuning or In-Context Learning (ICL) \cite{he2023chateda}, which are among the most promising methods for optimizing LLMs \cite{dai2022can}. AnalogCoder \cite{lai2024analogcoder} to our knowledge is among the first Analog circuit generator and generated the circuit through prompt engineering ICL. AmpAgent \cite{liu2024ampagent} is designed for multi-stage amplifier schematic design as well as process and performance porting.

\textbf{SPICE.} SPICE is a computer-based tool widely used by engineers for simulating and modeling electronic circuits. By performing mathematical analysis, it allows for the prediction of circuit behavior. SPICE can simulate a variety of components, from basic passive elements like resistors and capacitors to more advanced semiconductor devices like MOSFETs, making it essential for circuit design and optimization. Components are defined by their names, the nodes they are connected to, and their values, including resistors (R), capacitors (C), inductors (L), and transistors (M for MOSFETs, Q for BJTs). The netlist describes the interconnections of these components across the circuit's nodes.

\section{How tall does LLM stand in SPICE Code generation?} \label{howfar?}  
While LLMs demonstrate exceptional performance across various generative tasks, such as question answering, language translation, and conversational agents, these applications primarily involve natural language processing, an area in which LLMs receive extensive training. In contrast, their proficiency in managing specialized languages and tasks that are less frequently encountered during pretraining, such as generating SPICE code for hardware design, remains uncertain. Therefore, to effectively employ LLMs in automating hardware design tasks like SPICE code generation, it is essential to develop a comprehensive understanding of the capabilities and limitations of state-of-the-art LLMs. This understanding can prevent both undue optimism and unwarranted pessimism regarding their application. Our evaluation aims to provide this insight, establishing a foundation for future advancements in LLM-driven automated hardware design. To achieve this objective, we conducted an in-depth exploration and identified the common limitations of existing LLMs in the context of SPICE code generation. Ultimately, by addressing the identified shortcomings, we can reevaluate the potential of LLMs for practical automation in SPICE-based hardware design.

\subsection{Misconception of Gate Width and Length}
\begin{figure}[t]
    \centering
    \includegraphics[width=1.01\linewidth]{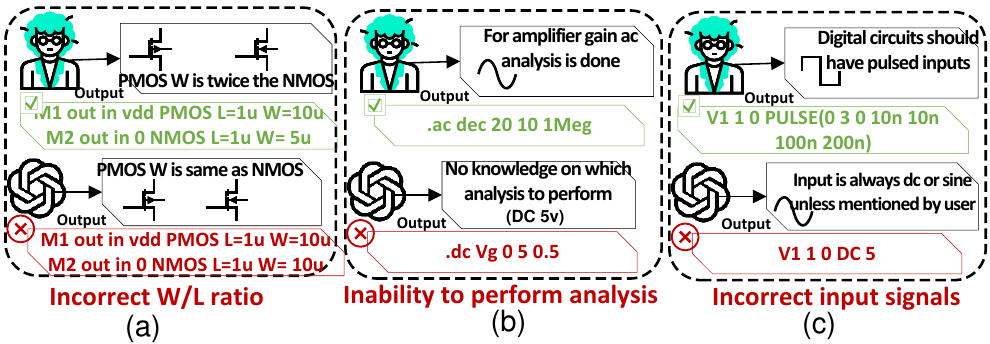}\vspace{-1.25em}
    \caption{Illustration of the errors generated by LLM for hardware in SPICE. (a) Incorrect W:L ratio, (b) Inability to perform circuit analysis, (c) incorrect input signals. \vspace{0.1em}}
    \label{errors}
\end{figure}

In SPICE code simulation, the precise definition and selection of gate lengths and widths are paramount to accurate circuit simulation and analysis. However, LLMs exhibit notable misconceptions in this area, and these misconceptions have a direct impact on circuit performance. Specifically, LLMs need a more foundational understanding of critical circuit design principles, such as the 2:1 PMOS to NMOS gate width ratio in many standard design practices to balance the drive strength (or current-carrying capability) between the two types of transistors, as NMOS transistors generally have higher electron mobility compared to the hole mobility in PMOS transistors. Fig. \ref{errors}(a) shows that a novice or expert in SPICE coding is aware of the 2:1 gate width ratio basics, and on the other hand, the LLM fails. Failure to respect design norms can result in improperly balanced circuits that deviate from intended performance characteristics. The inability of LLMs to recognize and apply these conventions suggests a gap in their comprehension of circuit domain knowledge. It concerns their efficacy in generating accurate SPICE codes for circuit designs. We reported the results of our zero-shot prompting to generate SPICE codes for basic digital circuits (i.e., inverter, NAND, NOR) in Table \ref{llms} analyzing LLMs' power in generating the correct W:L ratio. We observe that plain implementation of under-test LLMs fail.  

\begin{table}[b]\vspace{-1em}
\centering
\caption{Analysis of LLM's ability in SPICE generation.}\vspace{-4.65 pt}
\label{W/L comparision}
\scalebox{0.75}{\begin{tabular}{lcccccc}
\hline
\rowcolor[HTML]{C0C0C0} 
\cellcolor[HTML]{C0C0C0} & \multicolumn{2}{c}{\cellcolor[HTML]{C0C0C0}\textbf{W/L ratio}} & \multicolumn{2}{c}{\cellcolor[HTML]{C0C0C0}\textbf{Input signal}} & \multicolumn{2}{c}{\cellcolor[HTML]{C0C0C0}\textbf{Analysis}} \\ \cline{2-7} 
\rowcolor[HTML]{C0C0C0} 
\multirow{-2}{*}{\cellcolor[HTML]{C0C0C0}\textbf{LLM}} & \textbf{Plain} & \textbf{Pilot Prompt} & \textbf{Plain} & \textbf{Pilot Prompt} & \textbf{Plain} & \textbf{Pilot Prompt} \\ \hline
GPT-4o & \xmark & \cmark & \xmark & \cmark & \xmark & \cmark \\
GPT-3.5 & \xmark & \cmark & \xmark & \cmark & \xmark & \cmark \\
Gemini-Adv & \xmark & \cmark & \xmark & \cmark & \xmark & \cmark \\
Claude-3.5 sonnet & \xmark & \cmark & \xmark & \cmark & \xmark & \cmark \\
Codestral & \xmark & \cmark & \xmark & \cmark & \xmark & \cmark \\
Mistral Large 2 & \xmark & \cmark & \xmark & \cmark & \xmark & \cmark \\
Mistral Nemo & \xmark & \cmark & \xmark & \cmark & \xmark & \cmark \\ \hline
\end{tabular}}
\label{llms}
\end{table}

\subsection{Inability to Perform Circuit Analysis}
SPICE simulations are a powerful tool for various forms of circuit analysis, such as transient, DC, and AC analysis, each tailored to reveal specific characteristics of the circuit under study. However, a significant shortfall observed with LLMs, as shown in Fig. \ref{errors}(b), is their failure to autonomously select and perform the appropriate type of analysis based on the circuit's operational requirements. Effective circuit simulation often depends on the proper application of these analyses: transient analysis for time-domain response, DC analysis for steady-state behavior, and AC analysis for frequency response characteristics. LLMs have demonstrated an inability to grasp these distinctions, often leading to inappropriate or irrelevant analyses in response to user requests. This limitation highlights a deficiency in logical reasoning within the context of circuit simulation and an inadequate understanding of how these analyses contribute to evaluating circuit performance. Table \ref{llms} highlights that plain implementation of under-test LLMs fail to perform proper circuit analysis.

\subsection{Incorrect Input Signal Assignment and Device Parameter Configuration}
Accurate input signal assignment and device parameter configuration are essential for circuit testing within SPICE simulations. However, as shown in Fig. \ref{errors}(c), LLMs frequently fail to assign input signals in a manner consistent with the circuit's functional requirements, such as setting correct voltage levels, signal frequencies, or waveform types. Furthermore, fundamental parameters, including the device's operating temperature, should be considered or correctly set by LLMs. The temperature, which influences carrier mobility and, consequently, the overall behavior of semiconductor devices, must be carefully controlled to ensure realistic simulation results. The oversight of such basic parameters undermines the reliability of simulation outputs and reflects a need for a more nuanced understanding of device-specific considerations essential for precise SPICE simulations.


\section{SPICEPilot}\label{sec:framework}
In this section, we address \textit{RQ3}, \textit{RQ4}, and \textit{RQ5} posed in the introduction.
The proposed SPICEPilot framework focuses on leveraging the capabilities of LLM to generate hardware SPICE code within a Python environment using PySpice, while avoiding the costly process of fine-tuning. Given the scarcity of data, the goal is to enable the LLM to reason and creatively contribute to the development of intelligent models in circuit design. This is achieved by embedding fundamental circuit logic, reasoning, and error identification mechanisms. In the SPICEPilot framework, we introduce a computation-friendly method that leverages a reference containing step-by-step guidelines for code generation. This reference minimizes trivial errors and limits the need for extensive hardware knowledge, helping to avoid common mistakes that we discussed in Section \ref{howfar?}. The reference is used to validate every output generated by the LLM, significantly reducing errors.  This approach is akin to ICL\footnote{Dai et al. \cite{dai2022can} suggest that ICL can be viewed as implicit fine-tuning, where ICL produces meta-gradients through forward computation similarly to explicit fine-tuning, which updates model parameters through back-propagation.}. 

\begin{figure}[t]
    \centering
    \includegraphics[width=1\linewidth]{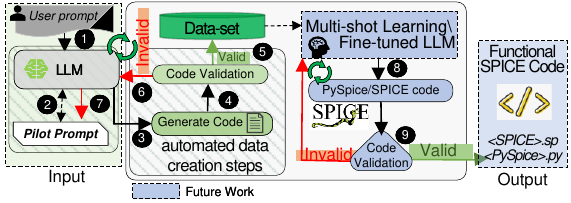} \vspace{-2em}
    \caption{SPICEPilot Framework.}
    \label{spicepilot}
\end{figure}

\subsection{Framework}
Fig.~\ref{spicepilot} illustrates the initial methodology designed for data augmentation in circuit generation using LLMs. The process begins with the \encircle{1} \textbf{User Input}, where the user provides specifications for the desired circuit. This input is then processed through the \textit{Pilot Prompt} (\encircle{2}), which integrates hardware knowledge to help the LLM mitigate common errors and offer insights into PySpice modules and coding styles. Leveraging this refined prompt, the LLM generates the corresponding PySpice code (\encircle{3}). The generated code undergoes a \textbf{Validation} process (\encircle{4}) to ensure the netlist correctness, where a human expert reviews and corrects trivial errors such as keyword mismatches in Python with PySpice. Successfully validated code is added to the \textbf{Dataset} (\encircle{5}). If the netlist is improperly constructed, the code is deemed invalid, and additional comments detailing the errors are sent back to the LLM (\encircle{6}), prompting a revision. This iterative cycle (\encircle{3}$\rightarrow$\encircle{4}$\rightarrow$\encircle{6}$\rightarrow$\encircle{3}) continues until valid code is produced (\encircle{5}). Concurrently, the identified errors contribute to the ongoing optimization of the \textit{Pilot Prompt} (\encircle{7}), which is updated either manually by a human expert or automatically via scripting based on the errors encountered. The valid PySpice in step \encircle{3} also allows us to generate the SPICE netlist, which is processed to extract key parameters such as the number of transistors after code verification step \encircle{4}. From the analysis, we obtain metrics such as gain while maintaining associated metadata as data points. The curated dataset will then be effectively utilized in our future works to enhance the LLM through ICL or fine-tuning, aiming to generate more robust designs in PySpice and SPICE (step \encircle{8}). The enhanced model continues to produce code that is subjected to functional validation, adhering to the \textit{valid} and \textit{invalid} classifications (step \encircle{9}). In cases of invalid code generation, the model receives specific error feedback to facilitate continuous improvement. The ultimate goal of this methodology is to iteratively refine the LLM’s capability to generate accurate and functional circuit designs based on user inputs, leveraging continuous validation and prompt optimization to enhance performance and reliability. Ultimately, the framework outputs the functional SPICE (.sp) and PySpice (.py) codes.

\subsection{Dataset Generation and Benchmarking}
The data points currently are stored with each point representing both the Pyspice model and its corresponding SPICE representation. This dual representation enables our dataset to be utilized to generate SPICE models using conventional methods or for rapid simulations within a Python environment.
To address {\textit{RQ6}}, we aim to establish a baseline evaluation for the community. Our selection criteria for benchmarking are based on the transistor count within the circuit. This benchmark consists of both Digital and Analog circuits as depicted in Table \ref{benchmark}. We classify circuits as follows: those with a transistor count of 10 or fewer are categorized as ``easy''; those ranging from 11 to 25 as ``medium''; circuits with 26 to 45 transistors are deemed ``hard''; and circuits exceeding 45 transistors are classified as ``extreme.'' This initial benchmarking framework can be further refined and expanded by considering additional factors, such as the number of nodes, which would enhance the understanding of circuit complexity and improve overall benchmarking for the community.

Previous studies in this domain, such as \cite{lai2024analogcoder}, have discussed benchmarking with a limited set of circuits. For our benchmark, we conducted a meticulous search and selected 60 unique circuits, which is 150\% larger than the Analogcoder benchmark~\cite{lai2024analogcoder}, over 7.5$\times$ the number of circuits included in the ChipChat benchmark~\cite{blocklove2023chip}, and offers 250\% more circuits compared to the VeriGen benchmark~\cite{thakur2023verigen}.

\begin{table}[t]
\caption{Benchmark descriptions include selected digital and analog circuit design tasks. The tasks are categorized by difficulty levels—\textcolor{green}{easy}, \textcolor{orange}{medium}, \textcolor{pink}{Hard}, and \textcolor{red}{Extreme} using different background colors for distinction. \vspace{-0.6em}}
\label{new_bench}
\centering
\scalebox{0.52}{\begin{tabular}{|c|c|c|l|c|c|c|l|}
\hline
\rowcolor[HTML]{EFEFEF} 
\multicolumn{1}{|l|}{\cellcolor[HTML]{EFEFEF}\textbf{ID}} & \textbf{Name}                                                                                              & \textbf{\#T}               & \multicolumn{1}{c|}{\cellcolor[HTML]{EFEFEF}\textbf{Explanation}}                          & \textbf{ID}                                       & \textbf{Name}                                                                                                  & \textbf{\#T}                                      & \multicolumn{1}{c|}{\cellcolor[HTML]{EFEFEF}\textbf{Explanation}}                                                                   \\ \hline
\cellcolor[HTML]{9AFF99}5                                 & \cellcolor[HTML]{9AFF99}SR Latch                                                                           & \cellcolor[HTML]{9AFF99}8  & \begin{tabular}[c]{@{}l@{}}Two CMOS NOR gates;\\  each NOR uses 4 T\end{tabular}           & \cellcolor[HTML]{FFCCC9}{\color[HTML]{333333} 3}  & \cellcolor[HTML]{FFCCC9}{\color[HTML]{333333} Operational Amplifier}                                           & \cellcolor[HTML]{FFCCC9}{\color[HTML]{333333} 30} & Common-source op-amp                                                                                                                \\ \hline
\cellcolor[HTML]{9AFF99}6                                 & \cellcolor[HTML]{9AFF99}Buffer                                                                             & \cellcolor[HTML]{9AFF99}4  & \begin{tabular}[c]{@{}l@{}}Two inverters in series; \\ each inverter uses 2 T\end{tabular} & \cellcolor[HTML]{FFCCC9}{\color[HTML]{333333} 9}  & \cellcolor[HTML]{FFCCC9}{\color[HTML]{333333} Voltage Regulator}                                               & \cellcolor[HTML]{FFCCC9}{\color[HTML]{333333} 25} & with Overcurrent Protection                                                                                                         \\ \hline
\cellcolor[HTML]{9AFF99}18                                & \cellcolor[HTML]{9AFF99}Half-Adder                                                                         & \cellcolor[HTML]{9AFF99}12 & \begin{tabular}[c]{@{}l@{}}XOR gate: 8 T, \\ AND gate: 4 T\end{tabular}                    & \cellcolor[HTML]{FFCCC9}{\color[HTML]{333333} 17} & \cellcolor[HTML]{FFCCC9}{\color[HTML]{333333} Switched-Capacitor Filter}                                       & \cellcolor[HTML]{FFCCC9}{\color[HTML]{333333} 45} & \begin{tabular}[c]{@{}l@{}}20 T Op-Amp\\ 12 T switches\\ 8 T Ctrl and clock\\ 5 T biasing\end{tabular}                              \\ \hline
\cellcolor[HTML]{FE996B}9                                 & \cellcolor[HTML]{FE996B}\begin{tabular}[c]{@{}c@{}}CMOS Multiplexer\\  (4:1)\end{tabular}                  & \cellcolor[HTML]{FE996B}20 & \begin{tabular}[c]{@{}l@{}}Transmission gates \\ and control logic\end{tabular}            & \cellcolor[HTML]{FE0000}7                         & \cellcolor[HTML]{FE0000}Delta-Sigma Modulator                                                                  & \cellcolor[HTML]{FE0000}60                        & \begin{tabular}[c]{@{}l@{}}20 T integrator\\ 10 T comparator \\ 15 T DAC\\ 10 T for clock\\ 5 T for biasing\end{tabular}            \\ \hline
\cellcolor[HTML]{FE996B}20                                & \cellcolor[HTML]{FE996B}\begin{tabular}[c]{@{}c@{}}Flash Analog-to-Digital\\  Converter (ADC)\end{tabular} & \cellcolor[HTML]{FE996B}35 & \begin{tabular}[c]{@{}l@{}}For 3-bit resolution; 7 \\ comparators at 5 T each\end{tabular} & \cellcolor[HTML]{FE0000}12                        & \cellcolor[HTML]{FE0000}4-bit Synchronous Counter                                                              & \cellcolor[HTML]{FE0000}88                        & 4 D-flip-flops × 22 T                                                                                                               \\ \hline
\cellcolor[HTML]{FE996B}29                                & \cellcolor[HTML]{FE996B}3-bit Ripple Carry Adder                                                           & \cellcolor[HTML]{FE996B}60 & \begin{tabular}[c]{@{}l@{}}3 full adders; \\ each uses 20 T\end{tabular}                   & \cellcolor[HTML]{FE0000}15                        & \cellcolor[HTML]{FE0000}\begin{tabular}[c]{@{}c@{}}Successive Approximation\\  Register (SAR) ADC\end{tabular} & \cellcolor[HTML]{FE0000}80                        & \begin{tabular}[c]{@{}l@{}}10 T differential compar.\\ 20 T resistor-ladder DAC\\ 30 T SAR Logic\\ 20 T Ctrl and clock\end{tabular} \\ \hline
\end{tabular}
}
 \label{benchmark}
\end{table}

\section{Experimental Analysis}
This section presents the SPICEPilot implementation and evaluation in two key aspects: $(i)$ Demonstrating the framework's ability to generate results that surpass current standards, and $(ii)$ Establishing a comprehensive benchmark for robust evaluation.
The above experiments are conducted to critically evaluate and gain insights necessary for establishing more concrete standardization for the LLMs. The identified capabilities and limitations are further discussed in Section~\ref{sec:Discussions}.
We extensively evaluate the capability of LLMs in circuit design, including CodeLlama-70B-Instruct \cite{40}, Wizardcoder-33B-V1.1 \cite{41}, Llama3-70B \cite{52}, GPT-3.5 \cite{42}, and GPT-4o. CodeLlama and WizardCoder are code generation LLMs, fine-tuned on Llama2 \cite{54} and StarCoder \cite{55}, respectively. Llama-3 is the newest open-source general LLM. WizardCoder and Llama-3 are LLMs that outperformed GPT-3.5 on the HumanEval \cite{56} coding tasks \cite{57}. 
We adopt the `Pass@k' metric \cite{58} (k=1, 5) as our main evaluation standard, a widely used approach in code generation tasks \cite{40,41}. This metric quantifies the proportion of correct generations within $k$ independent attempts, where higher values denote better performance. We conduct $n$ trials ($n\geq k$) and compute Pass@k using the formula $1 - \frac{\binom{n - c}{k}}{\binom{n}{k}}$, where $c$ denotes the number of successful attempts.


In our experiment, we utilize the setup illustrated in Fig.~\ref{spicepilot}, employing Claude-3.5 Sonnet as the backbone LLM. The \textit{Pilot prompt} is provided as an initial reference for learning. Subsequently, we prompt the LLM to generate solutions for each task in the Analogcoder Benchmark (ACB) \cite{lai2024analogcoder}. Since our framework emphasizes prompt engineering, we adapted the ACB prompts to include more detailed verbal descriptions of the circuit design, as outlined in Table \ref{new_bench}. Table~\ref{comp_A_C} demonstrates the superior performance of our framework, with notable improvement of 52.90\% in Pass@1 scores and improvement of 1.91\% at pass@5 and generating all 24 circuits in the benchmark, validating the efficacy of fine-tuning free approach with our prompt-engineered to augment the data in enhancing circuit design automation

\begin{table*}[t]
\centering
\caption{Performance comparison across SPICEPilot to ACB for different models.} \vspace{-1em}
\label{comp_A_C}
\scalebox{0.84}{
\begin{tabular}{c|cc|cc|cc|cc|cc|cc|cc|}
\cline{2-15}
                                             & \multicolumn{2}{c|}{\cellcolor[HTML]{C0C0C0}\textbf{CodeLlama-70B}} & \multicolumn{2}{c|}{\cellcolor[HTML]{C0C0C0}\textbf{WizardCoder-33B}} & \multicolumn{2}{c|}{\cellcolor[HTML]{C0C0C0}\textbf{Llama3-70B}} & \multicolumn{2}{c|}{\cellcolor[HTML]{C0C0C0}\textbf{GPT3.5}} & \multicolumn{2}{c|}{\cellcolor[HTML]{C0C0C0}\textbf{GPT4}} & \multicolumn{2}{c|}{\cellcolor[HTML]{C0C0C0}\textbf{AnalogCoder}} & \multicolumn{2}{c|}{\cellcolor[HTML]{C0C0C0}\textbf{SPICEPilot (w/ Claude)}} \\ \hline
\multicolumn{1}{|c|}{\textbf{Task Level}}    & \multicolumn{1}{c|}{Pass@1}                 & Pass@5                & \multicolumn{1}{c|}{Pass@1}                  & Pass@5                 & \multicolumn{1}{c|}{Pass@1}               & Pass@5               & \multicolumn{1}{c|}{Pass@1}             & Pass@5             & \multicolumn{1}{c|}{Pass@1}            & Pass@5            & \multicolumn{1}{c|}{Pass@1}                & Pass@5               & \multicolumn{1}{c|}{Pass@1}                     & Pass@5                     \\ \hline
\multicolumn{1}{|c|}{\textbf{Easy (1-8)}}    & \multicolumn{1}{c|}{9.15}                   & 36.7                  & \multicolumn{1}{c|}{21.2}                    & 50.7                   & \multicolumn{1}{c|}{75}                 & 92.5                 & \multicolumn{1}{c|}{54.6}               & 82.2               & \multicolumn{1}{c|}{100}                & 100              & \multicolumn{1}{c|}{100}                  & 100                 & \multicolumn{1}{c|}{\textbf{100}}                        & \textbf{100}                        \\ \hline
\multicolumn{1}{|c|}{\textbf{Medium (9-13)}} & \multicolumn{1}{c|}{0}                    & 0                   & \multicolumn{1}{c|}{0}                     & 0                    & \multicolumn{1}{c|}{16.7}                 & 20                 & \multicolumn{1}{c|}{15.3}               & 36.5                 & \multicolumn{1}{c|}{82.7}                & 91.5              & \multicolumn{1}{c|}{22.7}                  & 91.4                   & \multicolumn{1}{c|}{\textbf{91.7}}                        & \textbf{94.3}                        \\ \hline
\multicolumn{1}{|c|}{\textbf{Hard (14-24)}}  & \multicolumn{1}{c|}{0}                    & 0                   & \multicolumn{1}{c|}{0}                     & 0                    & \multicolumn{1}{c|}{0.3}                 & 3.03                 & \multicolumn{1}{c|}{0}                & 0               & \multicolumn{1}{c|}{7.9}              & 14.2              & \multicolumn{1}{c|}{33.9}                  & 60.3                 & \multicolumn{1}{c|}{\textbf{47.7}}                        & \textbf{62.3}                        \\ \hline
\multicolumn{1}{|c|}{\textbf{Avg}}           & \multicolumn{1}{c|}{3}                    & 12                  & \multicolumn{1}{c|}{7}                    & 16.9                   & \multicolumn{1}{c|}{30.6}                 & 38.5                 & \multicolumn{1}{c|}{23.3}               & 39.6               & \multicolumn{1}{c|}{63.5}              & 68.6              & \multicolumn{1}{c|}{52.2}                  & 83.9                 & \multicolumn{1}{c|}{{\textbf{79.8}}}                        & {\textbf{85.5}}                        \\ \hline
\multicolumn{1}{|c|}{\textbf{\# Solved}}      & \multicolumn{1}{c|}{7}                      & 7                     & \multicolumn{1}{c|}{7}                       & 7                      & \multicolumn{1}{c|}{11}                   & 11                   & \multicolumn{1}{c|}{6}                  & 6                  & \multicolumn{1}{c|}{10}                & 10                & \multicolumn{1}{c|}{20}                    & 20                   & \multicolumn{1}{c|}{\textbf{21}}                        & {\textbf{24}}                        \\ \hline
    \end{tabular}}
\end{table*}



In the second experiment, we employed two closed-source models, integrating the \textit{Pilot prompt} to infer circuits from their internal knowledge. We randomly selected 10 circuits from our versatile benchmarking in varying levels of complexity, ranging from simple to challenging, to evaluate the performance of the LLM. The results in Table~\ref{our_benchmark} indicate a high pass ratio, with the models successfully generating circuits with different transistor counts. The initial percentage for hard circuits in benchmarking~\ref{our_benchmark} is low, and our observation revealed it is due to various factors, such as module definition in Python, which is not supported by Spice, and the tuning of the circuit, which can be later resolved by either using Chain-of-Thought (CoT) or simple repetitive asking. Additionally, our framework automates waveform generation in PySpice and SPICE code generation, facilitating easy validation of the circuits. The model outputs are further subjected to verification, ensuring the correct creation of circuit netlists. It is important to note that numerous parameters must be fine-tuned for functional verification of analog circuits, including the adjustment of resistors and coupling capacitors, which play a critical role in analog domain performance.

\begin{table}[t]
\caption{SPICEPilot performance on our enhanced benchmark.} \vspace{-1em}
\label{our_benchmark}
\begin{tabular}{c|cc|cc|}
\cline{2-5}
\textbf{}                                  & \multicolumn{2}{c|}{\cellcolor[HTML]{C0C0C0}\textbf{SPICEPilot (w/ GPT-4o)}} & \multicolumn{2}{c|}{\cellcolor[HTML]{C0C0C0}\textbf{SPICEPilot (w/ Claude)}} \\ \hline
\multicolumn{1}{|c|}{\textbf{Task Level}}  & \multicolumn{1}{c|}{Pass@1}                     & Pass@3                     & \multicolumn{1}{c|}{Pass@1}                     & Pass@3                     \\ \hline
\multicolumn{1}{|c|}{\textbf{Easy (10)}}   & \multicolumn{1}{c|}{100.0}                      & 100.0                      & \multicolumn{1}{c|}{90.0}                       & 100.0                      \\ \hline
\multicolumn{1}{|c|}{\textbf{Medium (10)}} & \multicolumn{1}{c|}{90.0}                       & 100.0                      & \multicolumn{1}{c|}{80.0}                       & 100.0                      \\ \hline
\multicolumn{1}{|c|}{\textbf{Hard (5)}}         & \multicolumn{1}{c|}{20.0}                       & 40.0                      & \multicolumn{1}{c|}{20.0} & 60.0 \\ \hline
\multicolumn{1}{|c|}{\textbf{Avg}}         & \multicolumn{1}{c|}{80.0}                       & 88.0                      & \multicolumn{1}{c|}{72.0}                       & 92.0                      \\ \hline
\multicolumn{1}{|c|}{\textbf{\# Solved}}   & \multicolumn{1}{c|}{20}                         & 22                         & \multicolumn{1}{c|}{18}                         & 23                         \\ \hline
\end{tabular}
\end{table}

\section{Discussions and Future Works}\label{sec:Discussions}
The research delves into the functionality of LLMs, revealing that while the framework generates accurate netlists in our experiments, achieving functional efficiency and meeting key parameters such as gain requires further knowledge instillation. This need arises due to the LLM's limited circuit-specific intelligence. To address this, we plan to enhance the LLM by providing it with high-definition circuit knowledge, enabling it to better integrate and resonate with detailed circuit behavior and design requirements.
Figure~\ref{spicepilot} presents the proposed framework aimed at addressing the significant data bottleneck in circuit generation through automated data augmentation, extension of it's applicability. This framework offers the community an initial methodology to streamline data generation processes. The approach can be extended to leverage multi-modal LLM by incorporating circuit images as inputs. Utilizing Python packages, the framework can be enhanced to the drawing of circuit and waveform representations for a better decoding, thereby facilitating reasoning based on visual inputs. The use of Python also allows for the implementation of class functions to construct circuits that can be integrated into more extensive design projects seamlessly. Additionally, the SPICE code generation dataset can be enhanced by incorporating detailed descriptions, similar to the methodology proposed in \cite{zhang2024mgverilog}. This enhancement assists LLMs in generating more sophisticated and accurate SPICE models, thereby advancing the capabilities of SPICE generation in LLM's.\vspace{-0.25em}

\section{Conclusion}
This paper presents SPICEPilot, an innovative approach to bridging the gap between software automation and hardware design in the realm of analog and digital circuits. By leveraging LLMs and PySpice, SPICEPilot automates the generation of SPICE code and introduces a reliable framework for benchmarking circuit performance. Our evaluation of both open-source and proprietary LLMs underscores the current limitations and future potential of AI-driven code generation in hardware design. Moreover, our proposed framework offers a solution to data scarcity through the generation of open-source datasets, paving the way for further advancements in the field. This work lays the foundation for future research aimed at optimizing LLMs for analog circuit applications, accelerating innovation in circuit design and automation

\vspace{-0.25em}
\small\bibliographystyle{IEEEtran}
\bibliography{IEEEabrv,./Reference}\vspace{-2em}

\end{document}